\documentstyle[sprocl,epsfig]{article}

\begin{document}
\title{QUARK CONFINEMENT IN THE \\ ANALYTIC APPROACH TO QCD}
\author{A. V. Nesterenko}
\address{Physical Department, Moscow State University,
Moscow, 119899, Russia \\E-mail: nesterav@thsun1.jinr.ru}
\maketitle
     In the early 60's in the papers \cite{Redmond,BLS} the so-called
``Analytic approach'' to quantum field theory was proposed. Its basic idea
is the explicit imposition of the causality condition, which implies the
requirement of the analyticity in the $Q^2$ variable for the relevant
physical quantities. This approach was successfully applied to the
analytization of the perturbative series when calculating the QCD
observables \cite{ILS}. It leads to the elimination of the unphysical
singularities, to the higher loop correction stability and to a weak
scheme dependence. However, the $Q^2$-evolution of some QCD observables
(for instance, the structure function moments) is intimately tied with the
solution of the renormalization group (RG) equation. Our task here is to
involve the analytization procedure into the RG formalism more profoundly.

     Let us consider the RG equation of a quite general form for quantity
${\sf A}(Q^2)$. At the one-loop level this equation and its solution read
\begin{equation}
\label{StdRGEqn}
\frac{d\ln{\sf A}(Q^2)}{d\ln Q^2} = \gamma\,
\widetilde{\alpha}^{(1)}_{\mbox{\small s}}(Q^2)
\qquad \Longrightarrow \qquad
{\sf A}(Q^2) = {\sf A}(Q^2_0)
\left[\frac{\widetilde{\alpha}^{(1)}_{\mbox{\small s}}
(Q^2_0)}{\widetilde{\alpha}^{(1)}_{\mbox{\small s}}(Q^2)}\right]^{\gamma}.
\end{equation}
Here $\gamma$ is the corresponding anomalous dimension,
$\widetilde{\alpha}^{(1)}_{\mbox{\small s}}(Q^2) =
\ln^{-1}(Q^2/\Lambda^2)$ is the one loop perturbative running coupling.
Obviously, the solution of Eq.\ (\ref{StdRGEqn}) has unphysical
singularities in the physical region $Q^2>0$. However, in many interesting
cases the quantity ${\sf A}(Q^2)$ must have correct analytic properties in
the $Q^2$ variable (namely, there is the only cutoff $Q^2\le 0$). One can
demonstrate this proceeding only from the first principles.

     At first sight, we come to the contradiction. Indeed, the left-hand
side of the standard RG equation (\ref{StdRGEqn}) has no unphysical
singularities in the $Q^2>0$ region, while its right-hand side has
pole-type singularity at the point $Q^2=\Lambda^2$. The account of the
higher loop contributions just introduces the additional unphysical
singularities of the cut type in the physical region $Q^2>0$ and hence
does not solve the problem. But there is no real contradiction here. This
situation arose with the perturbative expansion of the anomalous dimension
in the right-hand side of the RG equation.

     In order to improve the situation, we propose to use the following
method~\cite{PRD}. Before solving the RG equation (\ref{StdRGEqn}) one
should analytize its right-hand side as a whole for restoring correct
analytic properties. This prescription leads to the analytized RG
equation, which, at the one-loop level, takes the form
\begin{equation}
\label{ARGEqn}
\frac{d\ln{\sf A}(Q^2)}{d\ln Q^2} = \gamma\,
\widetilde{\alpha}^{(1)}_{\mbox{\small an}}(Q^2)
\qquad \Longrightarrow \qquad
{\sf A}(Q^2)={\sf A}(Q^2_0)
\left[\frac{^{\mbox{\scriptsize N}}\widetilde{\alpha}^{(1)}_{\mbox{\small
an}}(Q^2_0)} {^{\mbox{\scriptsize
N}}\widetilde{\alpha}^{(1)}_{\mbox{\small an}}(Q^2)}\right]^{\gamma}.
\end{equation}
Here $\widetilde{\alpha}^{(1)}_{\mbox{\small an}}(Q^2) =
\ln^{-1}(Q^2/\Lambda^2) + (1-Q^2/\Lambda^2)^{-1}$ is the perturbative
running coupling analytized by making use of the Shirkov--Solovtsov
prescription \cite{ILS}. So, the solution of the analytized RG equation is
expressed in terms of the new analytic running coupling which at the
one-loop level has the form \cite{PRD}:
\begin{equation}
\label{NARCDef}
^{\mbox{\scriptsize N}}\alpha^{(1)}_{\mbox{\small an}}(Q^2) =
\frac{4\pi}{\beta_0}\frac{z-1}{z\ln z},\quad
z = \frac{Q^2}{\Lambda^2}.
\end{equation}
The distinctive feature of this running coupling, that plays the crucial
role in the framework of our consideration, is its infrared (IR)
enhancement. It is worth noting that such a behavior of the invariant
charge is in agreement with the Schwinger--Dyson equations (see discussion
in Ref.\ \cite{Alek}), and, as it will be demonstrated further, provides
the quark confinement {\it without invoking any additional assumption}.

     At the higher loop levels there is only the integral representation
for $^{\mbox{\scriptsize N}}\alpha_{\mbox{\small an}}(Q^2)$. Figure~1
shows the new analytic running coupling computed at the one-, two-, and
three-loop levels. The essential merits of the running coupling
$^{\mbox{\scriptsize N}}\alpha_{\mbox{\small an}}$ are the following (see
Ref.\ \cite{PRD} for the details): It has no unphysical singularities at
any loop level; It is stable with respect to both the higher loop
corrections and the scheme dependence; Its singularity at the point
$Q^2=0$ is of the universal type at any loop level; Its IR behavior is in
agreement with the Schwinger--Dyson equations.

\medskip

     Let us turn to obtaining the quark-antiquark potential. We proceed
from the standard expression \cite{Brambilla,Rich} for this potential in
terms of the running coupling~$\alpha(q^2)$,
\begin{equation}
\label{VrGen}
V(r) = -\frac{16\pi}{3} \int_{0}^{\infty}\frac{\alpha(q^2)}{q^2}
\frac{e^{i{\bf qr}}}{(2\pi)^3}\,d{\bf q}.
\end{equation}
For the construction of the new interquark potential $^{\mbox{\scriptsize
N}}V(r)$ we shall use the new analytic running coupling (\ref{NARCDef}).
Upon the integration over the angular variables and some calculations (see
Ref.\ \cite{PRD} for the details) one can present this potential at large
distances in the following way:
\begin{equation}
^{\mbox{\scriptsize N}}V(r) \simeq \frac{8\pi}{3\beta_0}\Lambda
\cdot\frac{R}{2 \ln R}, \quad R = \Lambda r, \quad R \to \infty.
\end{equation}
Thus the new analytic running coupling $^{\mbox{\scriptsize
N}}{\alpha}_{\mbox{\small an}}(q^2)$ (see Eq.\ (\ref{NARCDef})) leads to
the rising quark-antiquark potential $^{\mbox{\scriptsize N}}V(r)$ which
can, in principle, describe the quark confinement. At the same time the
behavior of the potential $^{\mbox{\scriptsize N}}V(r)$ when $r\to 0$ has
the standard form \cite{Rich} determined by the asymptotic freedom:
$^{\mbox{\scriptsize N}}V(r) \simeq \Lambda (R \ln R)^{-1}$. It can be
shown also that the obtained result is stable with respect to both the
higher loop corrections and the scheme dependence \cite{PRD}.

     For the practical use of the new potential it is worth obtaining a
simple explicit expression that approximates it sufficiently well. For
this purpose one can use, for instance, the approximating function $U(r)$
(see Ref.\ \cite{PRD}). Of course, this function is not the unique one.
Nevertheless, the comparison of $U(r)$ with the phenomenological Cornell
potential shows their almost complete coincidence (see Fig.\ 2). A rough
estimation of parameter $\Lambda$ in the course of this fitting gives
$\Lambda \simeq 500$~MeV. Estimation of $\Lambda$ from the gluon
condensate for the $^{\mbox{\scriptsize N}}{\alpha}_{\mbox{\small an}}$
has been performed recently and gives the close value, $\Lambda=530\pm
50$~MeV.

\medskip

     Thus, we infer that combination of the RG formalism with the Analytic
approach, in principle, enables one to obtain the confining
quark-antiquark potential.

\bigskip
\noindent
\begin{center}
\begin{tabular}{cc}
\parbox[t]{55mm}{
\centerline{\epsfig{file=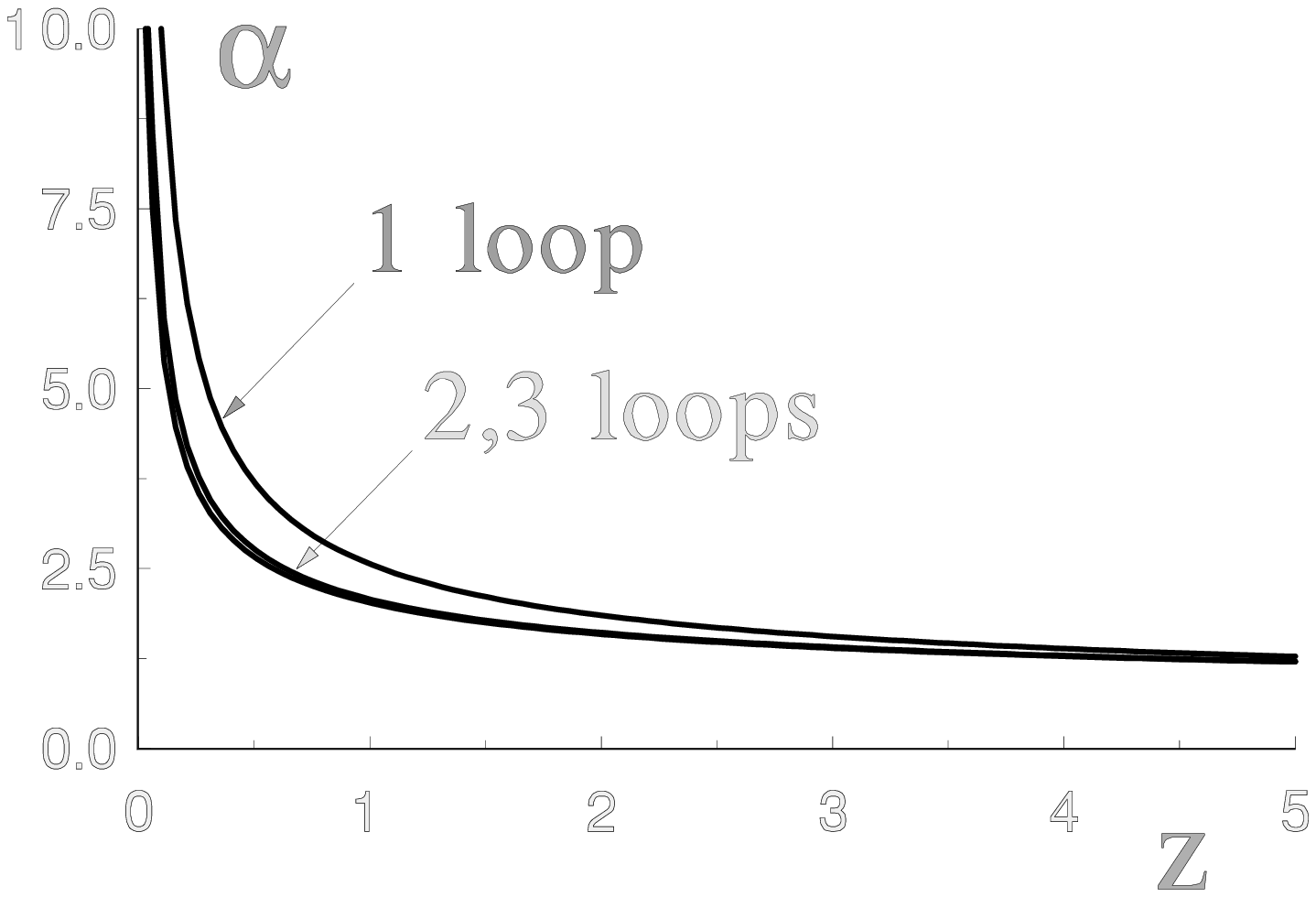, width=52.5mm}}
\parbox[t]{52.5mm}{\footnotesize Figure 1: The normalized new analytic
running coupling at the one-, two-, and three-loop levels.}
}
&
\parbox[t]{55mm}{
\centerline{\epsfig{file=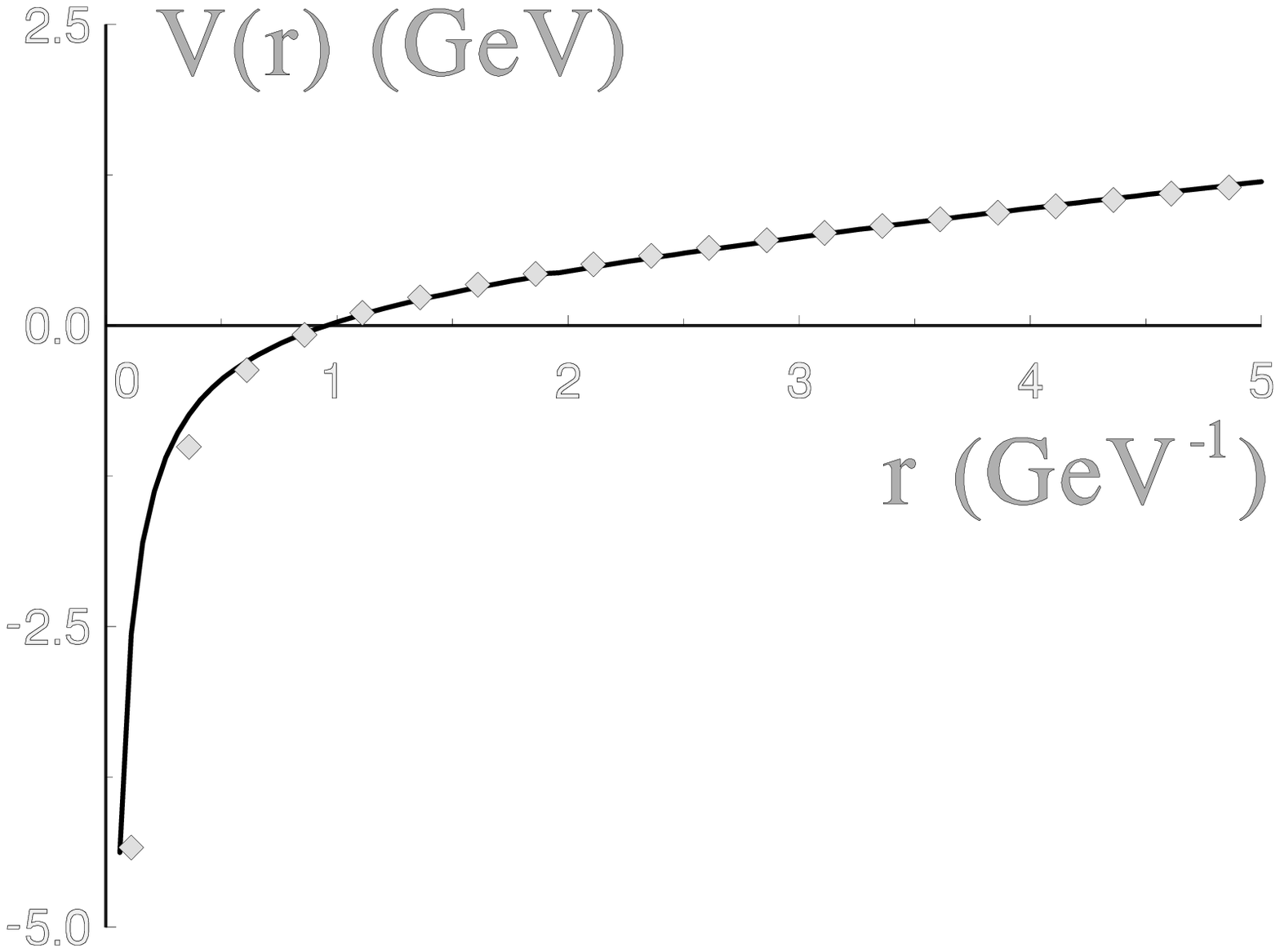, width=52.5mm}}
\parbox[t]{52.5mm}{\footnotesize Figure 2: Comparison of the $U(r)$
(solid curve) and the Cornell ($\Diamond$) potentials;
$\Lambda=530$~MeV, $n_f=5$.}
}
\end{tabular}
\end{center}

\end{document}